\documentclass{aa}  
\usepackage{txfonts}
\usepackage{amssymb}
\usepackage{hyperref}
\usepackage{color}
\usepackage{textcomp}
\usepackage{psfig}
\usepackage{amssymb}
\usepackage{epsfig,graphicx,natbib,url}

\usepackage{color}
\usepackage[normalem]{ulem}
\usepackage{enumitem}

\begin{document}

\title{Texture of average solar photospheric flows and the donut-like pattern}
\author{T. Roudier\inst{1}
\and
        J. Ballot\inst{1}
\and
       J.M.~Malherbe\inst{2}
\and
       M. Chane-Yook\inst{3}
}
\offprints{T. Roudier,\\
\email{thierry.roudier@irap.omp.eu}}

\institute
    {
      Institut de Recherche en Astrophysique et Plan\'etologie (IRAP), Universit\'e de Toulouse, CNRS, UPS,
      CNES, 14 avenue Edouard Belin, 31400 Toulouse, France
\and
Observatoire de Paris, LESIA, 5 place Janssen, 92195 Meudon, France, PSL Research University, CNRS, Sorbonne Universit\'es,
UPMC Univ. Paris 06, Univ. Paris Diderot, Sorbonne Paris Cit\'e 
\and
Institut d’Astrophysique Spatiale, CNRS, Université Paris-Saclay, 91405 Orsay, France
}

\date{Received march , 2022; Accepted xxx}

\abstract
{ Detailed knowledge of surface dynamics is one of the key points in understanding magnetic solar activity. The motions of the solar surface, to which we have direct access via the observations, tell us about the interaction between the emerging magnetic field and the turbulent fields.}
{ The flows computed with the coherent structure tracking (CST) technique on the whole surface of the Sun allow for the texture of the velocity modulus to be analyzed and for one to locate the largest horizontal flows and determine their organization. }
{ The velocity modulus maps show structures  more or less circular and closed
which are visible at all latitudes; here they are referred to as donuts. They reflect the most active convective cells associated with supergranulation. These annular flows are not necessarily joined as would seem to indicate the divergence maps. }
{The donuts have identical properties (amplitude, shape, inclination, etc. ) regardless of their position on the Sun. The average donuts computed from all the donuts shows an asymmetry east-west of the amplitude which is related to previous works on
the wave-like properties of supergranulation. A kinematic simulation of the donuts' outflow applied to passive scalar (corks) indicates 
the preponderant action of the selected donuts which are, from our analysis, one of the major actors for the magnetic field diffusion on the quiet Sun.}
{ The absence of donuts in the magnetized areas (plages) indicates the action of the magnetic field on the strongest supergranular flows and thus modifies the diffusion of the magnetic field in that location. The detection of the donuts is 
a way to locate -- in the quiet Sun -- the vortex and the link with the jet, blinkers, coronal bright points (campfires), or other physical structures.

Likewise, the study of the influence of donuts on the evolution of active events, 
such as the destruction of sunspots, filament eruptions, and their influences on upper layers via spicules and jets, could be done more efficiently via the detection of that structures.
}


\maketitle

\section{Introduction}

  Dynamics of the solar photosphere has a direct impact on solar magnetism and the evolution of higher layers of the atmosphere, that is, chromosphere and corona. More  specifically, the supergranular-scale flows on the solar surface contribute to the solar cycle by diffusing the magnetic flux in the quiet Sun and eroding active regions. Understanding the organization and the physical properties of motions at the solar surface is thus essential to constrain these processes. Different dynamical scales are observed in the kinetic energy spectrum  at the surface of the Sun, with the two main scales being the granulation ($\sim 1$\:Mm) and the supergranulation ($\sim 30$\:Mm). The physical nature of the latter is still elusive and needs to be better understood. Different approaches can be used to determine the properties of supergranulation: it can be seen either by directly following the dynamics of the photosphere thanks to Dopplergrams or reconstructions of horizontal flows, (local) helioseismology \citep{DG00,Ber2004,HGSD08,Lang2015}, or the organization of the magnetic field or its absence (voids), for example \citep{OBK2012,Berrilli2014}.

   The  lattice of the bright emission network  visible in the \ion{Ca}{ii} K line
   \citep{SL64} is located at the edge of the supergranules, and thus it is also a way to characterize the supergranulation. However, this approach may be biased by the way the network is drawn automatically (e.g., tesselation of the \ion{Ca}{ii}  network or segmentation of the divergence field or watershed basin lines  \citep{Srik2000,DET04,DBDK04,Gold2009}) or manually. It can be subject to  the law of closure (also known as the Gestalt laws of grouping), which states that breaks in an object are not  perceived and that the object is viewed as continuous, in a smooth pattern.
  So, \ion{Ca}{ii} patterns (or other proxies, such as the magnetic field, or UV AIA 1660 or 1700 \AA)  in some cells are well defined but many are not and they can be detected automatically by eye or using algorithms  without being sure of their validity.
  
 In Dopplergrams, it is difficult to find matching pairs of blue- and red-shifted features that define supergranules all over the solar surface, especially at the disk center \citep{Cliver2021}. Computing spatial spectra of flows is a common approach to identify the most energetic spatial scales of the photosphere. However, this method does not allow us to obtain detailed morphological and temporal properties of supergranules \citep{Getling2022}. In the same way, the works by  \citet{Svanda2013,LGB2015,LBG2016,LBG2018} used average supergranules to extract general properties, but this suppressed information on their localization and concealed their diversity in shape, size, amplitude, or orientation.
  The supergranulation properties are nevertheless  relatively sensitive to the technique used. 
  Thus the measurement of the characteristics of supergranules (shapes, sizes, lifetimes, evolution, and downflows) is a challenge, especially in the disk center.

Supergranules associated with the horizontally moving material, with a divergence  structure signature, are generally described as being distributed roughly uniformly over the entire solar surface.    More specifically, supergranulation characterized by radial flows is often assimilated as jointed cells of various sizes and shapes.  It has been noticed that observed properties could depend on the instrument used to quantify the supergranulation flows, with some consequences on the physical descriptions of supergranulation  \citep{Wil2014}.  The heterogeneity of flows in terms of size and amplitude  is well visible in Fig.~1 of \citet{RMRRBP08}. This large field of view, at the disk center, not only reveals zones with  large amplitude diverging flows, which are not necessarily adjacent, but also shows large regions with low amplitude velocity with low or not  diverging vectors. 
 
 So we proposed to develop a different approach that minimizes various flaws of previously cited methods. Our method help us to describe the texture of the velocity field at the solar surface.  The texture analysis returns information on the spatial arrangement of intensities, in all or parts of the image. Here, the texture of the velocity modulus reflects a general appearance  with certain parts having a signature arrangement (circle, disk). 
 
 The CST \citep{Roud2012,SRRBG13,RRPM13,Rincon2017,ROUD2018} allowed us to follow the horizontal velocities on  the whole surface of the Sun with very high spatial (2.5 Mm) and temporal resolution (30 minutes). The CST is a set of codes written in IDL and Fortran 90, computing the horizontal velocity field on the Sun surface by using solar granules as a tracer. With this technique,  we can  describe the structure of surface velocities  in detail and locate, for example, the largest horizontal flows and analyze how they are organized (cells, waves, etc.). We can then verify if these structures are or are not linked to strong positive divergences. Our main goal is to identify the regions where the supergranulation is the most effective and to determine if it can be related to other physical structures, such as vortexes, jets, blinkers, or coronal bright points (campfires). The diffusion of the magnetic flux tubes by the supergranule flows is studied through the evolution of corks, that is, passive scalars. 
 
  For the present study, we took advantage of  the long-term observations of Helioseismic and Magnetic Imager (HMI) instrument onboard the Solar Dynamics Observatory (SDO) spacecraft   and we applied our new approach based on the CST.  We investigated the modulus of the surface velocity to describe the texture of photospheric flows. In Sect. 2, we describe the data selection and our reduction pipeline to get the velocity field in spherical coordinates. In Sect. 3, we present the texture of the photospheric flows, and we discuss the detection of the annular structures flows, called donuts, in the velocity modulus field. We describe their properties and characteristics in Sect. 4. The modeling and the evolution of passive scalar (corks) are presented in Sect. 5. Finally, in Sect. 6, we point out a possible link between the density of donuts and the magnetic field amplitude, before concluding in Sect. 7.

\section{HMI data selection and reduction}\label{sec:red}

The HMI instrument aboard the SDO spacecraft \citep{Scherrer2012,Schou2012} provides uninterrupted observations over the entire solar disk.  The HMI intensity-Dopplergram images  observed at wavelengths around the 617.3 nm spectral line of neutral iron \ion{Fe}{i} and allow for the photosphere at a 45-second cadence and a pixel of 0.5\arcsec\ to
be studied. This allowed us to measure the photospheric velocity fields during six consecutive days with uniform observation sets. Using the HMI/SDO white-light data from 26 November to 1 December 2018, we derived horizontal velocity fields from image granulation tracking \citep{Roud2012}.    Two additional sequences were also used for the present study: the first one being on 14 January 2016 to study the supergranulation properties in active areas and the second one being on 28 January 2018 to detect potential links between supergranulation and the eruptive filament.

Different corrections have been applied in alignment, relative to resizement,  limbshift correction \citep{Rincon2017}. The differential rotation profile was adjusted from the raw Doppler data averaged over the 24 h of observations.
  For the derotation procedure, we used a rotation profile that took only large scales into account (supergranules were averaged in space and time):
\begin{equation}
     \Omega(\theta) = A + B \sin^2\theta + C \sin^4\theta,
 \end{equation}
where $A=2.864 \times 10^{-6}$, $B=-5.214 \times 10^{-7}$, and $C=-2.891 \times 10^{-7} \mathrm{rad\cdot s^{-1}}$.
        
It gives a velocity of 1.9934 km/s  (or 14.1781\textdegree/day) at the equator. Solar differential derotation has been applied on 26 to 28 November from the left to the right (east to west) and on 29 November to 1 december from the right to the left (west to east), since the derotation reference date is 29 November at 00:00 UTC.

 The “derotation” of the solar data has been applied to the original data (intensity and Doppler). This consisted in correcting the mean differential rotation of the Sun to bring back longitudes related to the solar surface at the same locations for each time deviation from the origin of the first HMI image. The reference day for the CST code  (see manual of the CST codes 
  \footnote[1]{\url{https://idoc.ias.u-psud.fr/system/files/user_guide_annex_version1.2_26mars2021.pdf}}) was taken at 00:00  UTC, 29 November 2018, which is the middle of our sequence.  In the same way, for both sequences on 14 January 2016 and 28 January 2018, the time reference of the derotation was taken at 00:00  UTC for each date.
  
 From a CST \citep[]{Roud2012} of photometric structures, such as granules, we derived the projection of the photospheric velocity field ($u_x$, $u_y$) onto the plane of the sky/CCD matrix.  The $u_x$ and $u_y$ (in km/s) were computed at a cadence of 30 min with a spatial window of seven pixels, equivalent to 3.5\arcsec, that is, around 2.5 Mm. 
   The CST velocities were complemented by HMI Dopplergrams to form full velocity vectors. The pipeline running at Institut de Recherche en Astrophysique \& Plan\'{e}tologie (IRAP) resulted in full-disk maps of the zonal, meridional, and radial components of the surface flow with a cadence of 30~min. It has been shown in previous papers \citep{Rincon2017,ROUD2018,Roud2012} that the CST allows the study of both small ($>2.5$Mm)  and large velocity scales at the solar surface.

The combination of  the results with Doppler data, which provides the out-of-plane component $u_z$ of the velocity field, allowed us to calculate the full vector velocity field at the surface in solar spherical coordinates $u_{r}$,  $u_{\theta}$, and $u_{\phi}$, where $\theta$ and $\phi$ denote the latitude and the longitude, respectively. We computed the 30-min-averaged horizontal surface velocity field $\vec{u}_h \equiv  (u_{\theta},u_{\phi})$  \citep{ROUD2018}.  We obtained 288 velocity maps spanning over the six days of observation (26 November to 1 December 2018). The  velocity modulus  is $u_h$ = $|\vec{u}_h|$ = $\sqrt{ u_{\theta}^{2}+u_{\phi}^{2}} $.  

\section{Structure of horizontal flows}\label{reduct}
 
 \subsection{Photospheric horizontal flows' texture}\label{reduc}

\begin{figure}[!htbp]
\includegraphics[width=\hsize]{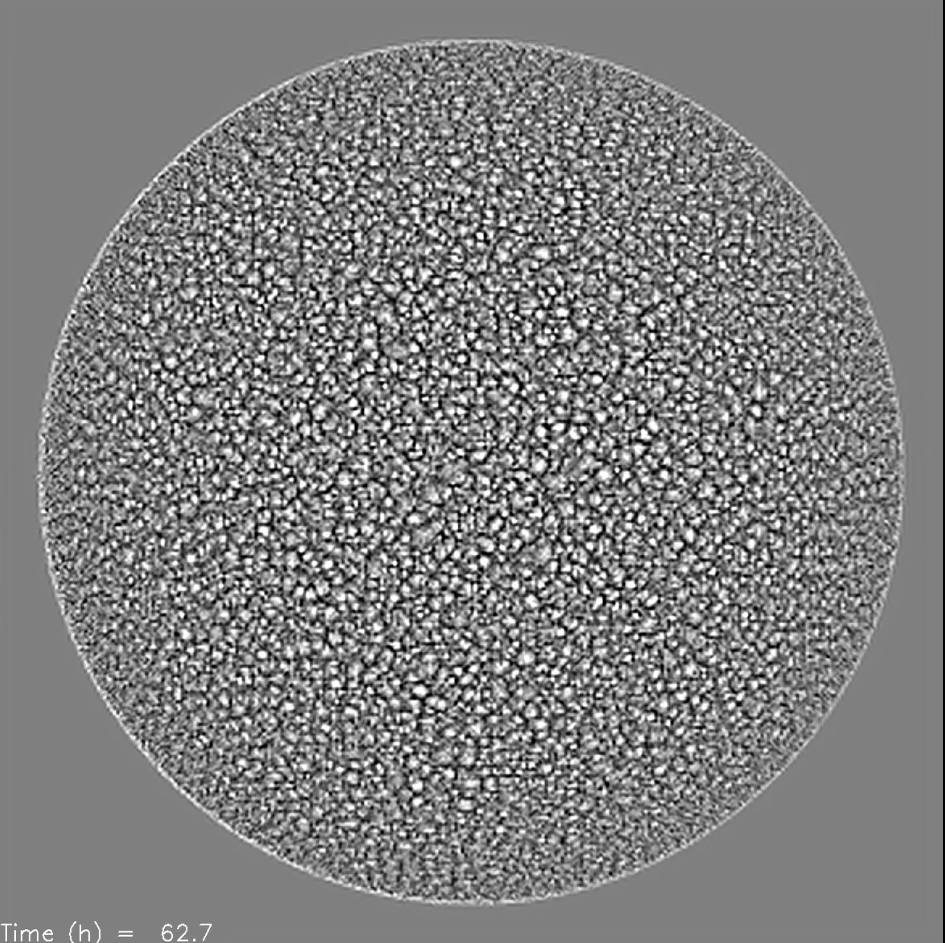}
\includegraphics[width=\hsize]{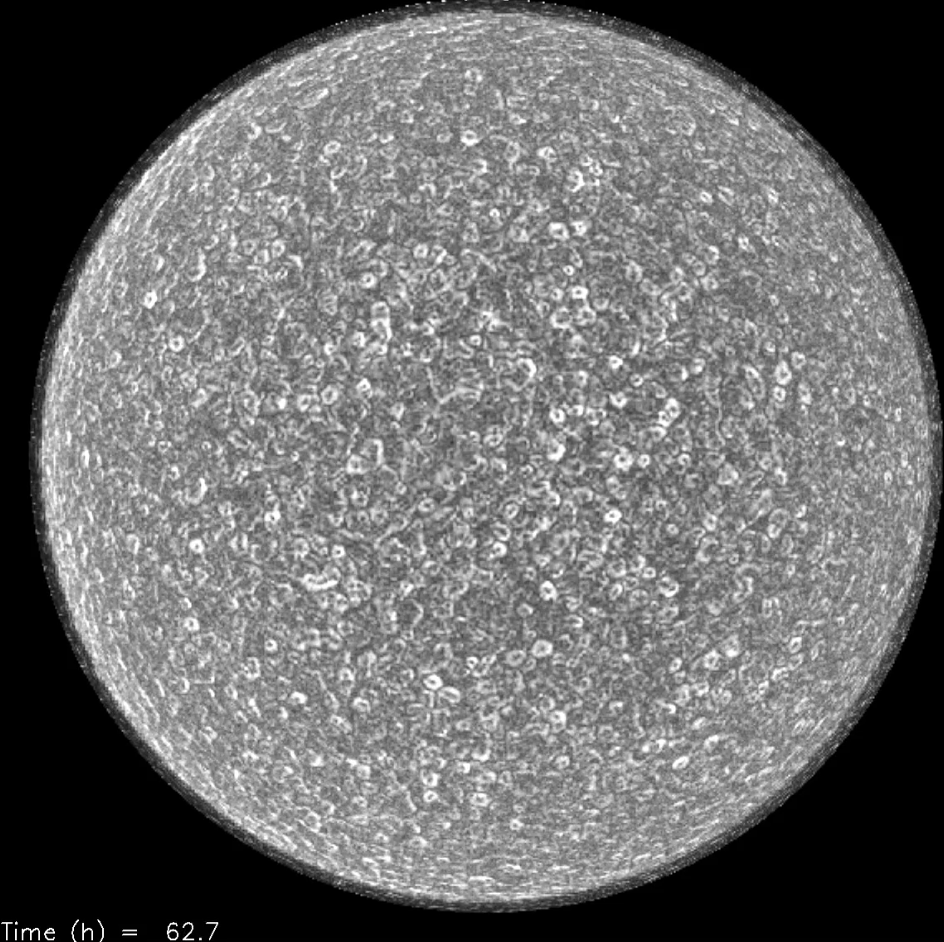}

\caption{ top) Divergence and (bottom) modulus maps of the horizontal velocity $\vec{u_{h}}$, averaged over 12\,hrs, on 29 November 2018.}
\label{uh}
\end{figure}

  A smoothing window of 12h was applied during the entire 6-day sequence to reduce the velocity noise and to study the texture,  in a topological context, of the photospheric horizontal flows $\vec{u_{h}}$  at the supergranulation timescale.
    The velocity modulus $u_h$ and horizontal divergence $\nabla\cdot\vec{u_{h}}$  were derived for every single snapshot of the 30-min horizontal velocity. An example horizontal divergence and the velocity modulus averaged over 12\,hr from 29 November 2018 are plotted in Fig.~\ref{uh}. The temporal evolution of the derotated  horizontal velocity modulus and divergence fields during 6 days can be watched  in movies 1 and 2 (Fig.~\ref{fig:movie1}, Fig.~\ref{fig:movie2}), respectively. A combination of movies 1 and 2 is given in movie 3 (Fig.~\ref{fig:movie3}).

 \begin{figure}[!htbp]
\includegraphics[width=\hsize]{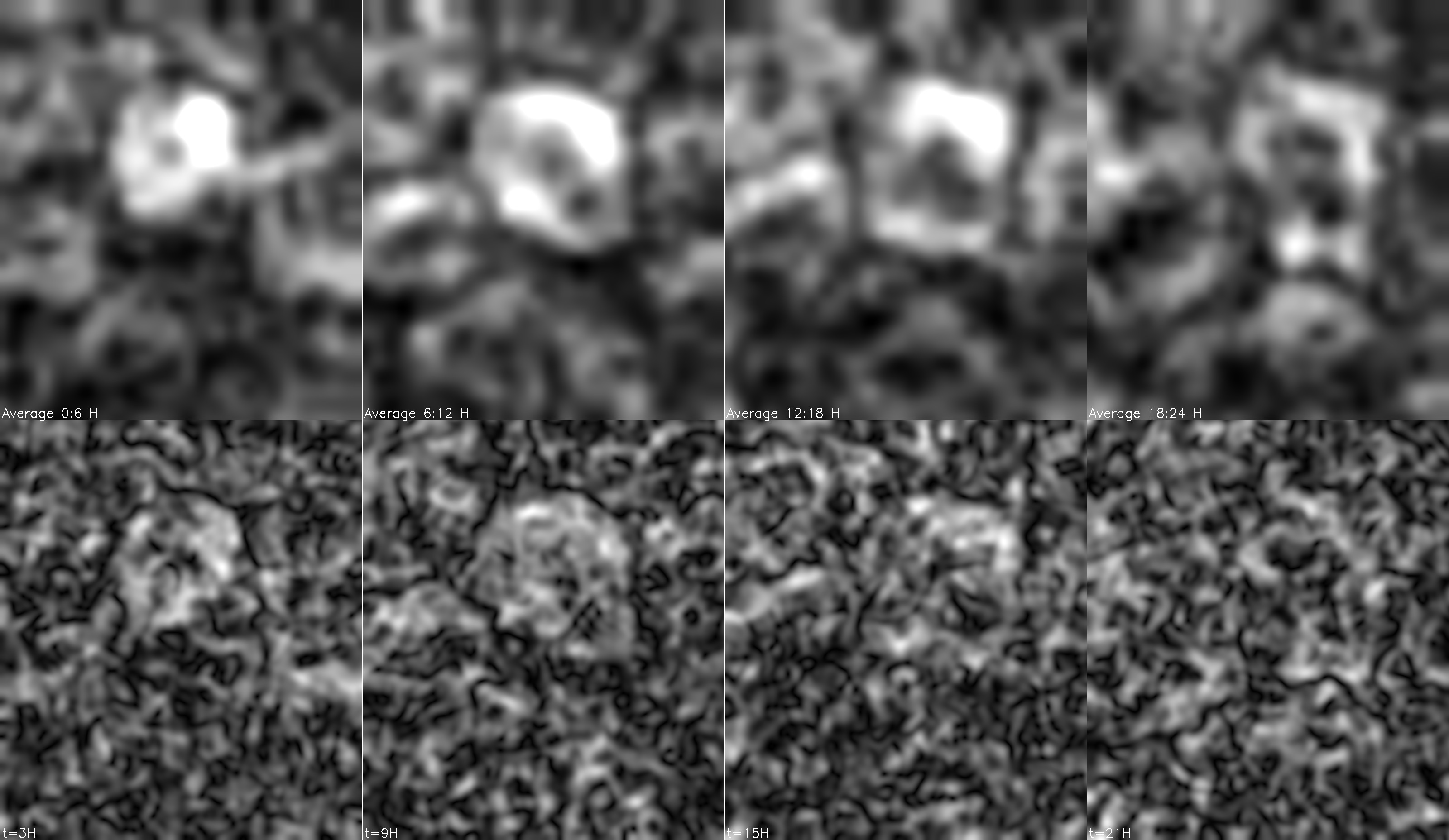}
\caption{Top row shows the evolution  of the horizontal flow modulus observed by Hinode from 29 August 2007, 10:17 UT to 31 August 2007, 10:19 UT, with a spatial resolution degraded to the one HMI/SDO. In each frame, the flow modulus is averaged over 6\,hrs. Each frame is separated by a time step $\delta t=6$\,h. The bottom images represent the 30-min averaged horizontal flow modulus corresponding to the middle time of each averaged top modulus maps. Fields of view span $90\arcsec \times 93\arcsec$ ($62.25\times67.43$ Mm$^2$). }
\label{hinode}
\end{figure}

 The divergence field shows positive structures (white) in Fig.~\ref{uh} (top), with a paving aspect over the entire solar surface, with different sizes and amplitudes. The maximum amplitude of that divergence structures was found to be  $9. 10^{-5}$\,s$^{-1}$ and  the mean size  corresponds to the supergranule one (about 20 Mm).  However, a region with positive divergence does not necessarily correspond to an isotropic structure in the underlying velocity field. Indeed, we observed that large amplitudes of the divergence can also be produced by a velocity gradient  in a privileged direction.

 The velocity modulus $u_h$, in Fig.~\ref{uh} (bottom), reveals another texture of the flow where the spatial organization of the pavement is less visible and is less distributed homogeneously with large areas of lower amplitude.
This kind of donut is also visible (see Fig.~\ref{hinode}) in a previous observation at a higher spatial resolution obtained with the Hinode/SOT instrument  \citep{RRBRM2009}  from 29 to 31 August 2007. Figure~\ref{hinode} shows a clear donut that is visible  in the horizontal velocity modulus. Its  amplitude is greater than the one of its neighborhood. The larger amplitude of velocity is located close to the edge of  the tree of fragmenting granules (TFGs). The top and bottom images of   Fig.~\ref{hinode} show the effect of the temporal averaging, which increases the isotropic aspect of the donuts.  One possible scenario of donuts' creation is due to the horizontal velocities' evolution  related to the TFGs \citep{Roudier2016}.  These  coherent flows tend to form the isotropic flows (donuts)  from the development of spatially close and temporally, almost simultaneous, TFGs \citep{Roudier2016}.  The other locations with large divergences, but weak unstructured $u_h$, are probably generated by isolated TFGs, with a lower velocity amplitude and none being in phase with their neighbors.
The most striking patterns in this map are the almost circular and closed structures visible at all the latitudes. These  structures, with an annular-shaped flow that formed by the 12h average of the horizontal velocity magnitude, certainly reflects the most active convective cells associated with supergranulation.  These  annular-shaped flows are also visible with a 2h average, but they are noisier. This irregular diffuse rings have been observed  previuosly by \citet{shine1999}. For convenience, we referred to these annular structures as “donuts” in that paper (and not in solar 
 physics in general) because these elements represent only a characteristic of the motions of the surface at the scale of the supergranulation.  The present works are mainly focused  on the characterization of these elements. 

On the full Sun velocity modulus map, large areas without a donut are visible. In these regions, the amplitude of the flows are weak and they are not clearly structured.

This kind of donut is also visible (see Fig.~\ref{hinode}) in a previous observation at a higher spatial resolution obtained with the Hinode/SOT instrument  \citep{RRBRM2009}  from 29 to 31 August 2007. Figure~\ref{hinode} shows a clear donut that is visible  in the horizontal velocity modulus. Its  amplitude is greater than the one of its neighborhood. The larger amplitude of velocity is located close to the edge of  the tree of fragmenting granules (TFGs). The top and bottom images of   Fig.~\ref{hinode} show the effect of the temporal averaging, which increases the isotropic aspect of the donuts.  One possible scenario of donuts' creation is due to the horizontal velocities' evolution  related to the TFGs \citep{Roudier2016}.  These  coherent flows tend to form the isotropic flows (donuts)  from the development of spatially close and temporally, almost simultaneous, TFGs \citep{Roudier2016}.  The other locations with large divergences, but weak unstructured $u_h$, are probably generated by isolated TFGs, with a lower velocity amplitude and none being in phase with their neighbors.

\subsection{Donut identification}\label{donuts}
   
          We aim to characterize the donuts in terms of locations, sizes, amplitudes, lifetimes, and shapes. At the beginning of this study, different techniques to isolate the donuts automatically were performed, such as watershed segmentation, neural networks' processing, among others. However, the results of these various approaches did not allow us --  in our special case of donut selection -- to isolate, exhaustively with confidence, all the visually detected donuts. This is especially due to the low-velocity contrast, the residual noise, and the inhomogeneous intensity repartition in the donuts  that were easily identified  by eye and brain of observers, and to the small number of structures (110) in the data to educate, for example, a neural network.
       
        Then, to start our study on a greater number of donuts, we decided not  to   use automatic recognition of the donuts and detected them directly by eye by checking them attentively.   In that first step, we concentrated our work on the two central days (28 and 29 November 2018) where the derotation allowed for the full Sun to be covered. 
          This enlarged donut sample was used, in a second step (see chapter 6.1), to detect the donuts as automatically as possible with  confidence.

        In order to flawlessly detect all the structures with an annular shape, we used a widget to watch the temporal evolution in all  locations on the Sun surface. In the present study, we are interested in isolating the largest amplitude and clearly recognizable $u_h$  annular structures. Some low amplitude and not fully closed structures with a short lifetime are probably lost; however, regarding our goal, we consider that  their contributions are small, in particular in the magnetic element diffusion for example.

       During that period of two days, 554 donuts were detected on the area located between $\pm60^\circ$ in latitude and longitude, giving a rate of donuts of $1.05\times 10^{-9}$ Mm$^{-2}$\,s$^{-1}$. Applying the impact assessment reasoning proposed by  \citet{TTTFS89} for exploding granules to our donut detection, we found, for an average donut area of 353 Mm$^2$, an impact of a donut of 747 hrs (i.e., 31 days) for each point on the Sun.

      \subsection{Donuts' lifetimes and location}\label{donuts1}

\begin{figure*}[!htp]
\includegraphics[width=\hsize]{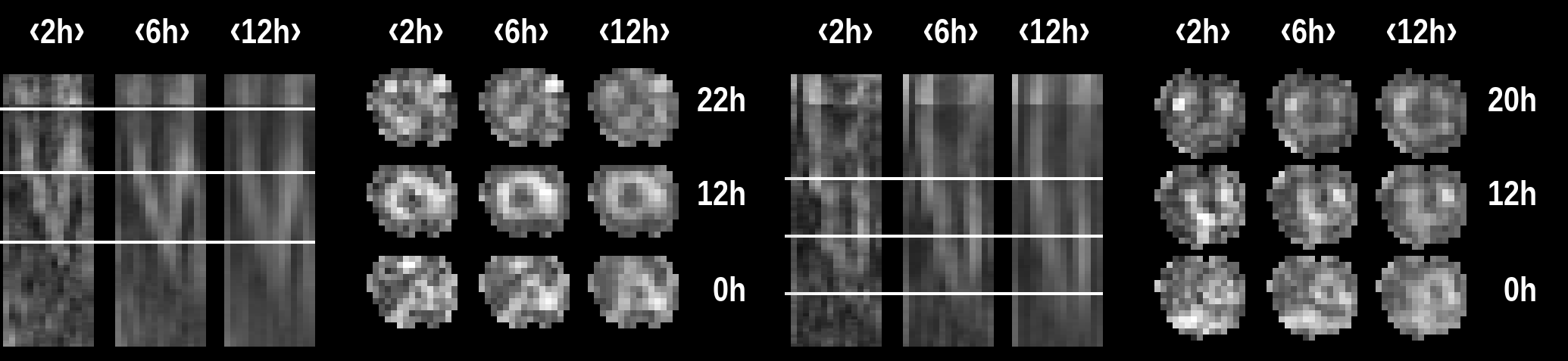}
\caption{  Two examples of donut evolution from their birth (a central hole appears) to their death (broken structure). Temporal cuts with different temporal averaging windows (2h, 6h, and 12h) in the center of each donut (on the left of each example) are shown.  On the right side, for each donut, cuts are shown at different times (0h, 12h, and 22h left donut example) and (0h, 12h, and 20h for the right donut example). These times are overplotted for each donut in their temporal respective cuts. In each case, the birth and death times correspond to the low and top white line of the respective temporal cuts. Plotted fields span $8\arcsec \times 8\arcsec$ ($5.8\times5.8$ Mm$^2$).}
\label{fig:donuts99}
\end{figure*}

       To measure their lifetimes, the data from 27 and 30 November 2018 were used to determine their births and deaths times.
        Due to the solar derotation, the edge of the solar disk was too noisy and only donuts located between $\pm60^\circ$ in latitude and longitude were studied.  The amplitude of the velocity modulus $u_h$ issued from the spherical components $u_{\theta}$ and $u_{\phi}$ does not suffer from projection effects; only the shape is affected (circle to ellipse) and it has been taken into account as for area measurements. The donut birth detection is a delicate point, particularly due to our visual procedure; even the donuts grow up with a ring shape from the beginning. We chose to define the birth of the donuts when the central hole appears in the circular shape of the structure. Due to the low spatial resolution of our data, the determination of the date of birth is defined at $\pm 1$\,h.  The date of death of the donuts was determined when the circular shape is broken. To limit the effect of the temporal window, the donuts' lifetimes were estimated with a temporal window of 2\,h.  We did not consider the evolution of donut pieces resulting from donut breaking and their possible impact on the surface dynamics. However, we roughly estimated that those pieces typically survive only one or two hours. Thus, regardless of whether this extra time is taken into account, it would not significantly modify the lifetimes ranging between  2~h and 55~h with a maximum of around 24~h (see Fig.~\ref{fig:context} top left),   with a global incertitude of around 3~h, which is characteristic of the supergranule lifetime \citep{Rincon2018}. Figure~\ref{fig:donuts99} shows the temporal evolution, with different temporal windows  of two donuts, where the birth (low line of the temporal cut) and death (top line in the temporal cut) times are indicated.

 \begin{figure}[!htbp]
 \includegraphics[width=\hsize]{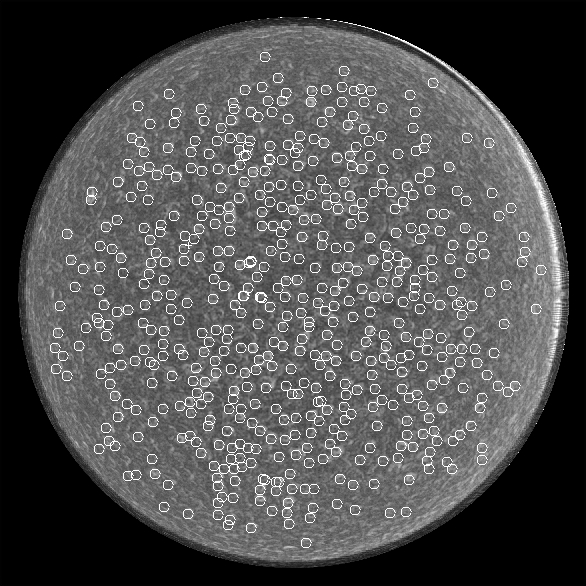}
\caption{Circles indicate the 554 detected donuts on a map showing the averaged horizontal velocity $u_{h}$ on 28-29 November 2018.}
\label{fig:donuts1}
\end{figure}

 \begin{figure}[!htbp]
 \includegraphics[width=\hsize]{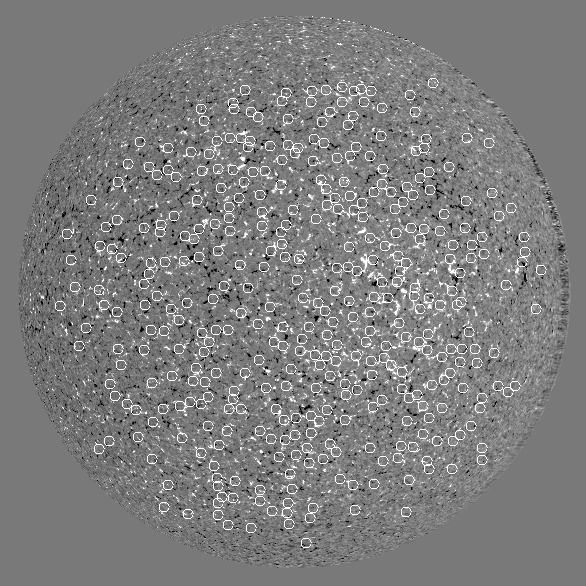}
\caption{Location of detected donuts (circles) superimposed on a map of mean longitudinal magnetic field $\langle B_{//}\rangle$  on 29 November 2018, averaged between 12:00 to 24:00 UTC.}
\label{fig:donuts2}
\end{figure}

 The location on the Sun surface of the 554 donuts visually detected between 28 and 29 November 2018 are superimposed (circle) on the average velocity modulus in Figure~\ref{fig:donuts1}.  It corresponds to a quiet Sun period.  We also compare the location of donuts with the magnetic field in Fig.~\ref{fig:donuts2}. Most  of the donuts are located in low amplitude magnetic field regions. A magnetic network is generally organized around the donuts. However, the links between the
 magnetic field and donuts  are  subtler than their relative location and this is described below (Sect.~5.3). The location of the donuts to the coronal hole visible in the SDO (193\AA) observation on 28 and 29 November 2018 does not reveal any particular spatial distribution of the donuts  relatively to the coronal hole
 \footnote[2]{\url{https://sdo.gsfc.nasa.gov/data/dailymov/movie.php?q=20181128_1024_0193}}.

\begin{figure}[!htp]
    \includegraphics[width=0.5\textwidth,clip=]{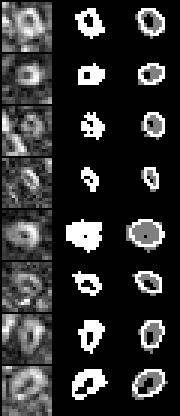}
    \caption{Examples of donuts' $ |$\vec{u_{h}}$| $  velocity modulus (left), segmented (middle) and adjusted with ellipses (right). The field of view is $58.9\arcsec \times 58.9\arcsec$ ($42.1\times42.1$\,Mm$^2$).
    \label{fig:donuts41}}
\end{figure}

\section{Donut properties and characteristics}\label{properties0}
        
         \subsection{Donut morphological properties}\label{properties1}

Once we detected donuts in time (birth to death) and space (latitude and longitude), we measured their shape and intensity distribution. More precisely, to determine the zone of influence of each  donut,  we averaged $u_h$  over its lifetime, and we applied a watershed technique to segment the mean structure \citep{Roudier2020}.  Figure~\ref{fig:donuts41} shows some examples of donuts averaged during their lifetime (left column) and their segmented (binarized) counterpart (middle column). To determine their shape, area, and orientation, each averaged donut was fitted with an ellipse (see Fig.~\ref{fig:donuts41}, right column). In addition, we measured their intensity distribution and evolution during their lifetime.

 \begin{figure}[!htbp]
\includegraphics[width=\hsize]{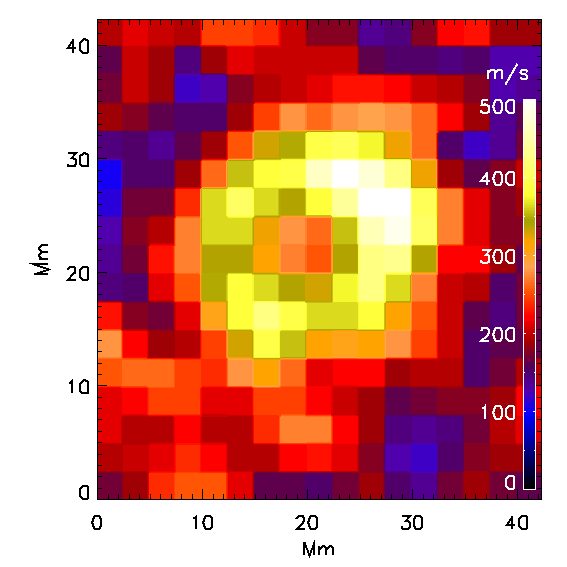}
\caption{Mean donuts computed from donuts in latitude and longitude $\pm$ 10\textdegree\ around the disk center. The estimated radius is 11 Mm with a mean modulus velocity at the center of $404\pm 78$~m/s and of $650\pm 65$~m/s  on the donut corona (white). }
\label{fig:donuts5}
\end{figure}

   \subsection{Mean donuts at the disk center}\label{properties1}
        First, we computed a mean donut close to the disk center ($\pm$ 10\textdegree) to avoid projection effects, by averaging all the 36 detected donuts  in that area. Figure~\ref{fig:donuts5} clearly shows the roundish shape of the “mean donuts” with an estimated radius of 11 Mm and a velocity amplitude, at the donut radius, up to 650 m/s. The modulus velocity in the center of the mean donuts of $404\pm 78$\,m/s reflects the diversity of the shape, intensity, and centering of all the averaged donuts. At the individual  donuts' center, at the disk center, the velocity is found to be around 157 m/s and is close to the level of the noise. An asymmetry of the velocity amplitude is clearly visible between the east (left) and west (right) sides. We could think that the differential rotation may produce such an asymmetry. However, as mentionned in Sect.~\ref{sec:red}, we applied a derotation procedure.   The observed excess of velocity amplitude on the west side, on the average donuts, is  quite probably of a solar origin. 
         This east-west anisotropy can be related to the previous observations showing that  asymmetry  in intensity \citep{LBG2016},  and related to anisotropy in the network where the network magnetic field is stronger in the west (in the prograde direction) than in the east \citep{LGB2015,Roudier2016}. This east-west anisotropy  is related to the wave-like properties of supergranulation. These waves were found to travel predominantly in the east-west direction, with more power in the prograde direction, and this excess in westward power leads to the observed super-rotation of the pattern \citep{LBG2018}.
     
          By widening the fields in latitude and longitude,  for example by considering a field of $\pm$ 20\textdegree\ with 130 detected donuts, we still observed this asymmetry of the velocity modulus of the mean donut. This suggests  that such a property is found up to the whole Sun and not only at the disk center.

    \subsection{Donut characteristics}\label{properties2}
 
 \begin{figure*}[!htp]
 \centerline{
   \includegraphics[width=0.45\textwidth,clip=]{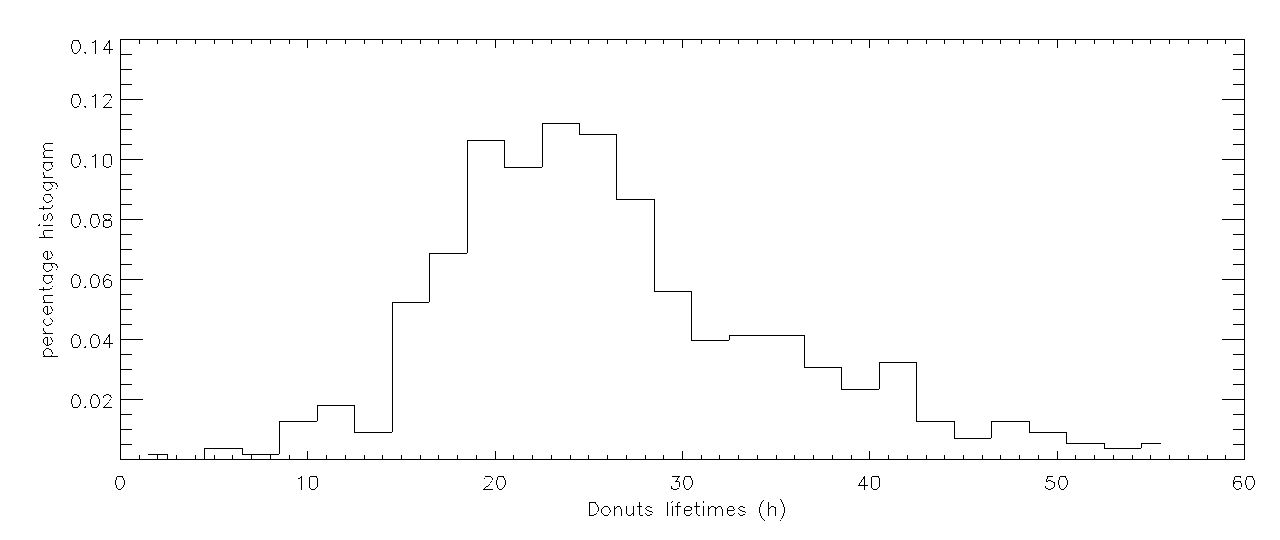}
   \includegraphics[width=0.45\textwidth,clip=]{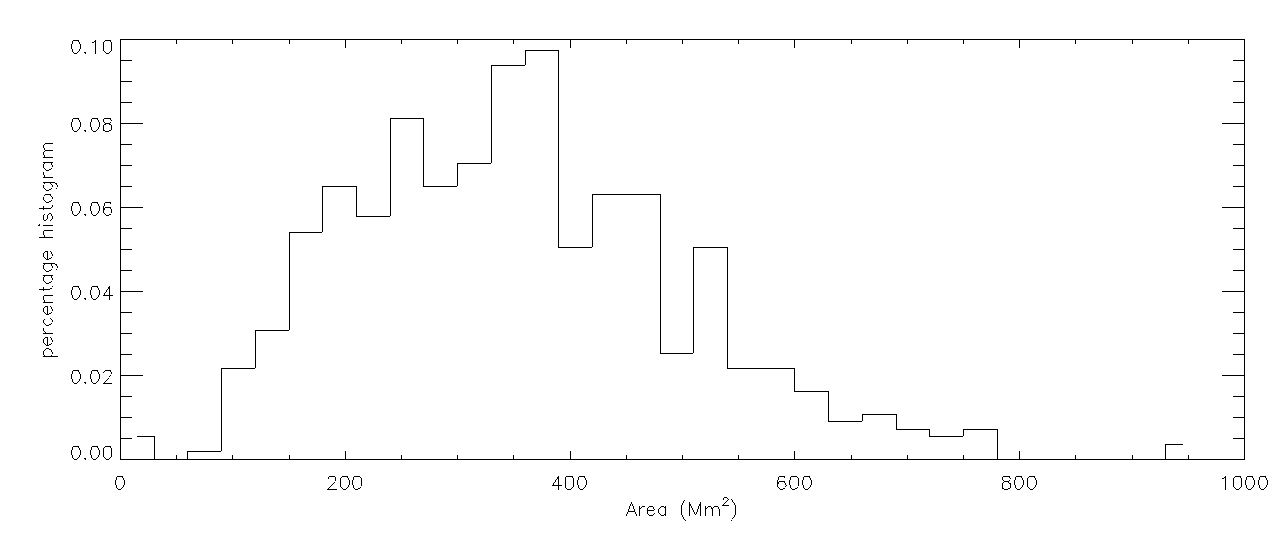}}
   
 \centerline{
  \includegraphics[width=0.45\textwidth,clip=]{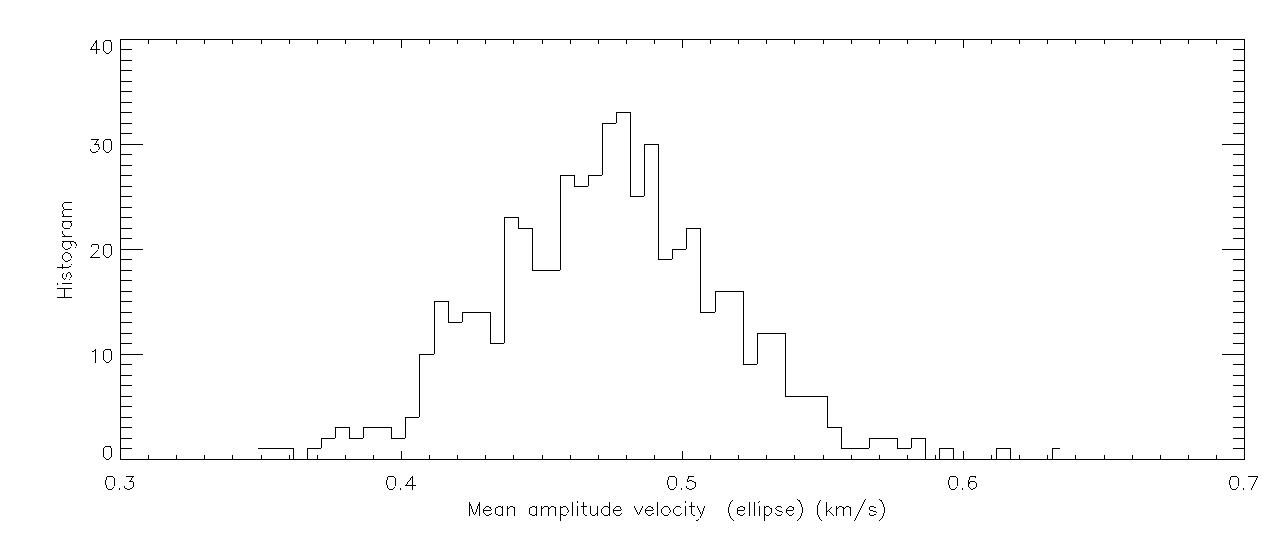}
  \includegraphics[width=0.45\textwidth,clip=]{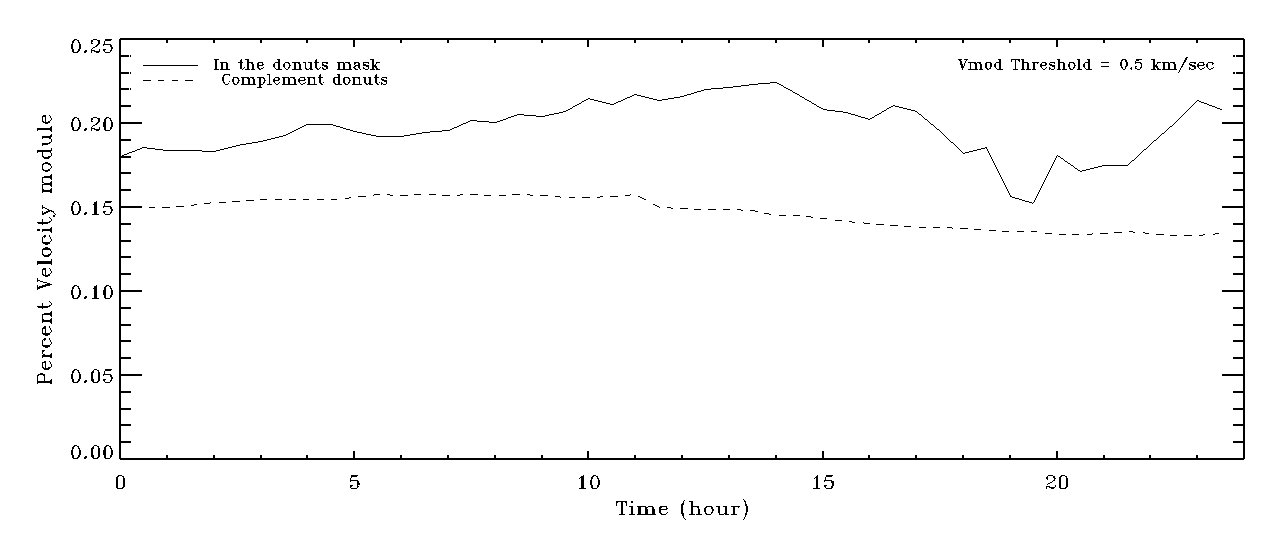}}
     
 \caption  { Top: Distribution of donut lifetimes (left) and distribution of donut areas corrected for projection effects (right).
    Bottom: Distribution of mean velocities inside donuts (left) and percentage of the area with a velocity greater than 0.5 km/s in donuts (solid line) and in the complementary field (dashed line) at the disk center $\pm20^\circ$ in latitude and longitude (right). These distributions come from data from 29 November 2018.}
 \label{fig:context}
\end{figure*}

     One of the advantages of the CST is to get the modulus surface flow $u_h$  at the disk center and then to  give the possibility to follow the donuts  with a high temporal resolution (2~h).  
     The segmentation of the donuts described above allowed us to determine the parameters that characterize these objects. 
       We measured their area and corrected them for spherical projection effects. 
     Their distribution is shown in Figure~\ref{fig:context} (top right). It lies between 80 and 900 Mm$^{2}$ and peaks around 350 Mm$^{2}$; this is equivalent to a radius of 10.6 Mm, which corresponds to the typical supergranule size.
 Surprisingly, no clear correlation has been found between the lifetime and the area: some small donuts have a long lifetime, whereas some large ones have short lives.

 In the same way, we did not find any correlation between the different parameters, such as the donuts corrected area, tilt angle, or the isotropy  of the donuts, and their location on the solar surface. This seems to indicate a relative homogeneity of the donut geometrical properties
 all over the solar surface.

The mean velocity modulus was extracted for each donut from its location and segmented area.  The mean velocity distribution peaks at 0.48 km/s (Fig.~\ref{fig:context} bottom left); this  does indeed correspond to the mean supergranular horizontal velocity amplitude \citep{Rincon2017}. The amplitude lies between 0.35 and 0.63 km/s. No correlation has been found between the mean velocity modulus and the donuts areas, and  also latitude,  confirming our previous remarks about the homogeneity of donut properties over the solar surface. 
We compared  the distribution of $u_h$ inside the donuts to its distribution in the full field at the disk center (in a field of $\pm20^\circ$ in latitude and longitude), on 29 November 2018. Values inside the donuts are significantly higher regardless of the temporal window used (see Table 1). This indicates that the detected donuts are the most important patterns in terms of horizontal flows on the Sun surface. This property is confirmed by the time evolution  of the percentage of area occupied by the velocity modulus higher than 0.5 km/s on the same field and in the donuts (Fig.~\ref{fig:context} bottom right).

\begin{table}[!htbp]
    \caption{ Mean velocity inside and outside donuts above different  velocity thresholds (0.5, 0.8,1.0 km/s)  at the location of $\pm 20$\textdegree\ in  latitude and longitude.
    The left column gives the temporal window used, the central column  corresponds to the mean velocity inside the donuts, and the right column is the  mean velocity outside the donuts }
    \label{tab:average}
    \centering
    \begin{tabular}{l r r}
          temporal &   mean velocity   &   mean velocity \\
          window   &    inside (km/s)  &   outside (km/s) \\
        \hline\hline
         \multicolumn{3}{c}{ only above 0.5 km/s} \\
        \hline
          2h  &   0.641  & 0.611  \\
          6h  &   0.626  & 0.590  \\
         12h  &   0.619  & 0.571  \\
         \hline\hline
         \multicolumn{3}{c}{  only above 0.8 km/s} \\
         2h. &   0.977   &  0.868  \\
         6h  &   0.970   &  0.842  \\
        12h  &   0.958   &  0.841  \\
        \hline\hline
     \multicolumn{3}{c}{ only above 1.0 km/s} \\
        \hline
          2h  &  1.196  &   1.040  \\
          6h  &  1.153  &   1.152  \\
         12h  &  1.147  &   1.033   \\
        \hline
    \end{tabular}
\end{table}

\begin{figure}[!htbp]
\includegraphics[width=\hsize]{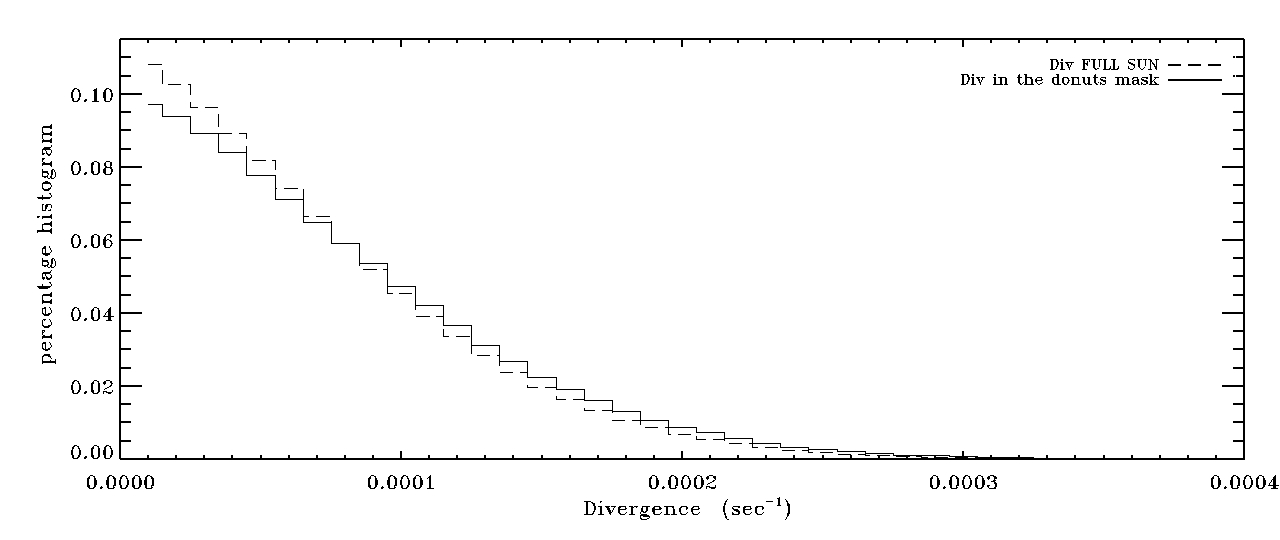}
\caption{Distribution of divergence amplitude in donuts (solid line) and in the full Sun (dashed line).}
\label{fig:donuts22}
\end{figure}

  We also explored the difference between the amplitude of the divergence inside and outside the donuts. We show their distributions in Fig.~\ref{fig:donuts22}. Comparing both histograms reveals that the divergence has a larger amplitude inside the donuts. Thus, the donuts represent the most powerful divergent events  on the solar surface.

\section{Impact of donut evolution on surface diffusion}\label{ssec:density}

  From our previous analysis, the donuts reflect the most active convective cells associated with supergranulation. 
  So it appears natural to measure the influence of these photospheric motions, linked to donuts, on the evolution of solar magnetism and activity. The transport of those magnetic elements in the photosphere and understanding the mechanism that diffuses magnetic fields is
still a challenge. Using the proxy of magnetic elements, we can follow the evolution of an initially randomly distributed passive scalar
(corks) to characterize the transport properties of the turbulent velocity field of the solar surface.

\subsection{Donut modeling}\label{ssec:model}

We aim to determine the influence of donuts on the evolution of the photospheric magnetic field on the solar surface. To avoid the other components of the velocity field (solar rotation, meridian circulation, etc.) and to isolate the role of donuts, we built the velocity field generated by only the simulated radial outflows of the donuts.

\begin{figure*}[!htbp]
\includegraphics[width=\hsize]{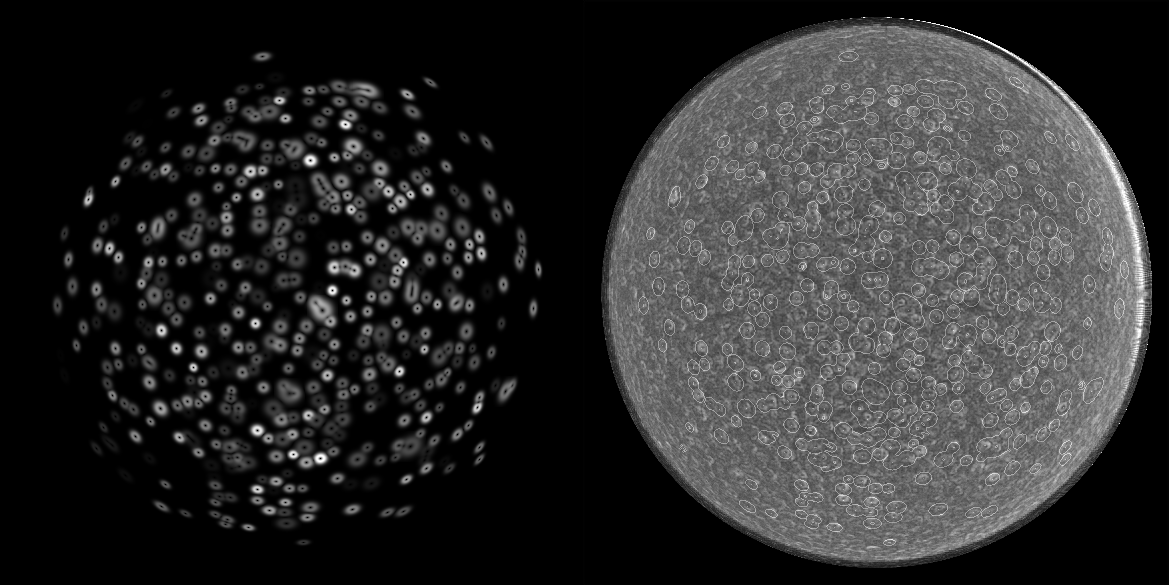}
\caption{ { Left: Map of $u_h$ of simulated donuts averaged on 28-29 November 2018, taking into
account their locations, lifetimes, velocities amplitude, and the projection effects. 
Right: Observed velocity modulus $ u_h$  averaged over 28-29 November 2018 and contour of the simulated donuts overlapped on the same period.}}
\label{fig:donuts34}
\end{figure*}

 To simulate the horizontal flows generated by donuts, we used a simple kinematic description, previously used by \citet{SW89,STW91}. Thus, we modeled the horizontal velocity inside a donut as a divergent flow around its center with a modulus
 
 \begin{equation}
     u_h(r)= u_0 \frac{r}{R} \exp \left[-\frac{1}{2}\left(\frac{r}{R}\right)^2\right],
 \end{equation}
 where $r$ is the distance to the center, $R$ the radius of the donut, and $u_0$ the amplitude of the field velocity. This profile was obtained by assuming that the source of the horizontal flow follows a Gaussian distribution.
  
  We note that this function is purely qualitative; the divergence of $u_h$ is zero for $r = R$, positive 
  for $r < R$, and negative for $r > R$.  It assumes circular symmetry, which is a first approximation.
   Such a spatial isotropic function was used to model  each of the 554 donuts, as well as the   3D files taking into account how the spatial and temporal location of them is built.
   The temporal average of this simulation is illustrated in Fig.~\ref{fig:donuts34} (left) and can be compared to the averaged velocity modulus on 28--29 November 2018 in Fig.~\ref{fig:donuts34} (right). To get a more realistic modeling of the donut outflow in function of their location on the solar surface,  we took into account the projection effects with the center-to-limb angle. 
    Some donuts were not included in the simulation because they were born just before the end of the 2 days and they have been excluded. 
   In the velocity histogram of all the simulated donuts' flows,  most of the velocity modulus lies between 0 and 400\,m/s, which is  perfectly representative of the amplitude supergranulation horizontal flows.   
 
  \subsection{Evolution of corks on the solar surface }\label{ssec:model2}

\begin{figure}[!htbp]
\includegraphics[width=\hsize]{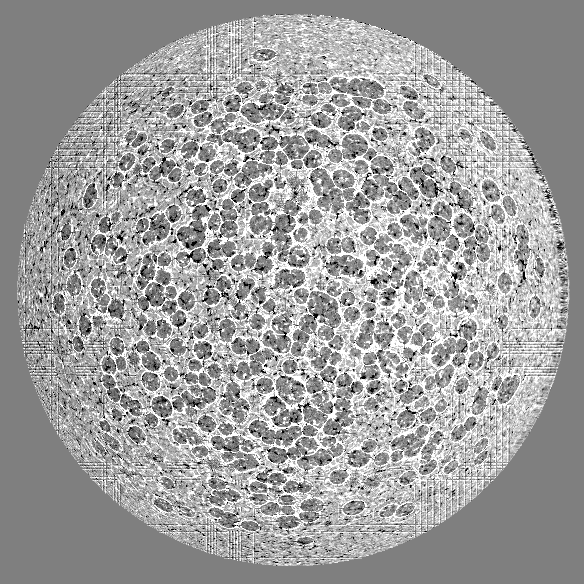}
\caption{ Final positions of the corks after two days of evolution overplotted on the mean magnetic longitudinal field $\langle B_{//} \rangle$  on 29 November 2018, from 12:00 to 24:00.}
\label{fig:donuts29}
\end{figure}

    To mimic the behavior of the magnetic flux tubes' evolution  on the solar surface, \citep{STW91} we used  passive test particles (also called corks) whose trajectories are guided by the flow created by the donuts.
The flow displaces the corks from  their initially uniform grid distribution.  The development of cork patterns in the field, over the two days, evolved from the uniform distribution over the surface to a structured shape.  Figure~\ref{fig:donuts29} shows the final positions of all the corks relative to the averaged longitudinal magnetic field between  12:00 and 24:00 on 29  November 2018. The fraction of the area devoid of corks was estimated at $65\%$ in the disk center  (to avoid projection effects). The network created by the diffusion of these corks is not exactly superimposed on the magnetic network of the quiet Sun, but it shows a large number of magnetic empty regions of these free zones of corks.  This seems to us to be a consequence of the selection of axisymmetric structures (donuts) that we have carried out,  losing the other surface motions that do not fit in the donut category. Nevertheless, we  see the influence of the only selected structures (donuts) which have a preponderant action in the diffusion of the magnetic field in the quiet Sun. The selected donuts are one of the major actors in the magnetic field diffusion on the quiet Sun.  This suggests that the distribution of the magnetic field on the solar surface depends on the most energetic donuts to form the photospheric network or magnetic patches. Small donuts, which do not take our selection into account, seem to play a minor role in the network building.

      We took advantage of the multiwavelength observations to compare the positions of the donuts detected on 29\ November 2018, with the solar corona observations being made on the same day. Where coronal holes are present, we observe that the regions are emptied of their corks, but it is not clear whether this is statistically significant when compared to the far 
      regions on the solar surface. No formal conclusion can be drawn as to the influence of the donuts' surface distribution and the coronal hole from our sequence.

\subsection{Regions with a low velocity modulus}\label{ssec:particularity}
 
 \begin{figure}[!htbp]
\includegraphics[width=\hsize]{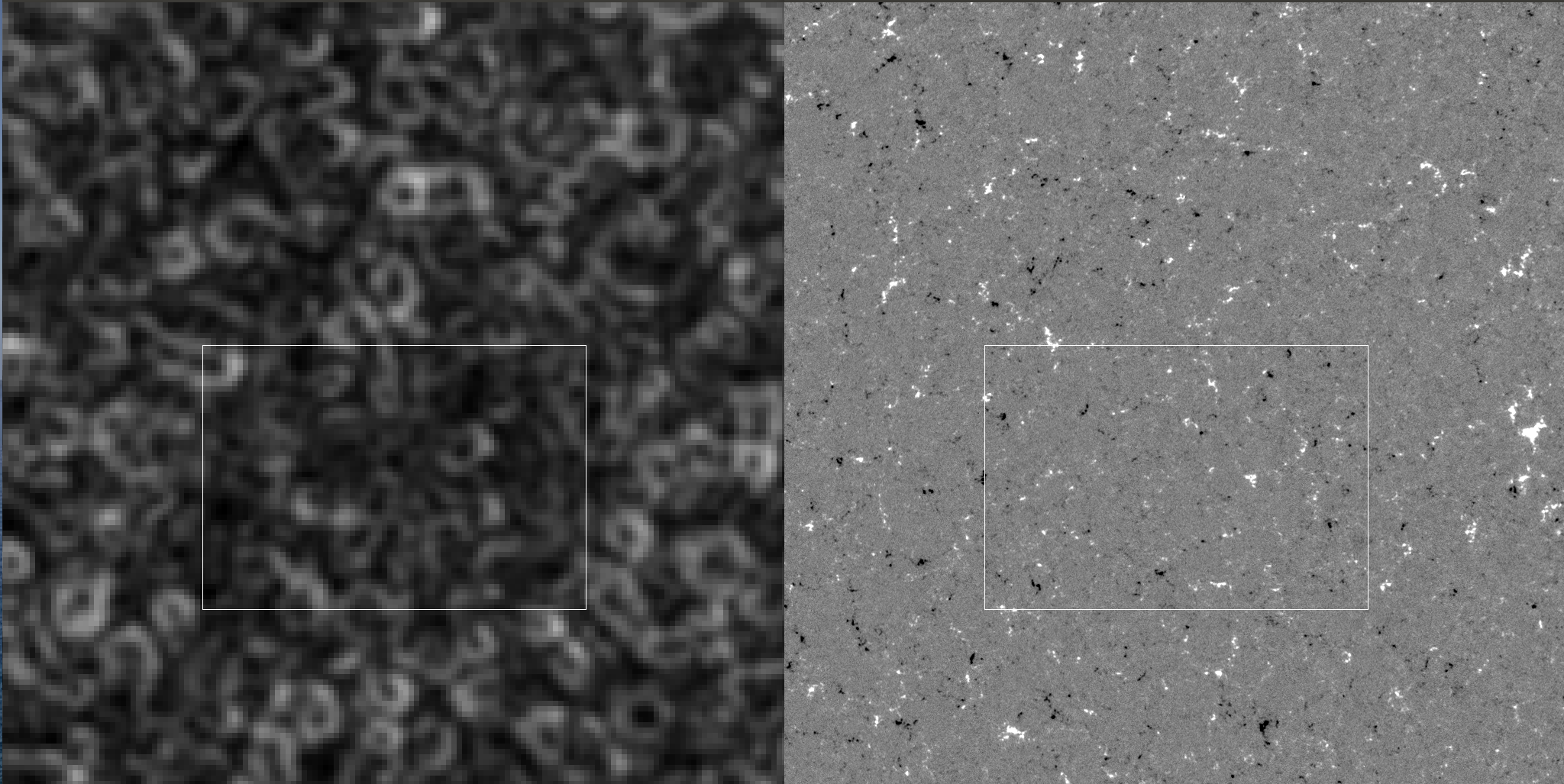}
\caption{Low velocity modulus region is in the white rectangle (field of view:  $257\arcsec \times 180\arcsec$ ($184\times 120 Mm^2$)) as well as the corresponding longitudinal magnetic field (B//).}
\label{fig:donuts39}
\end{figure}

  \begin{figure}[!htbp]
\includegraphics[width=\hsize]{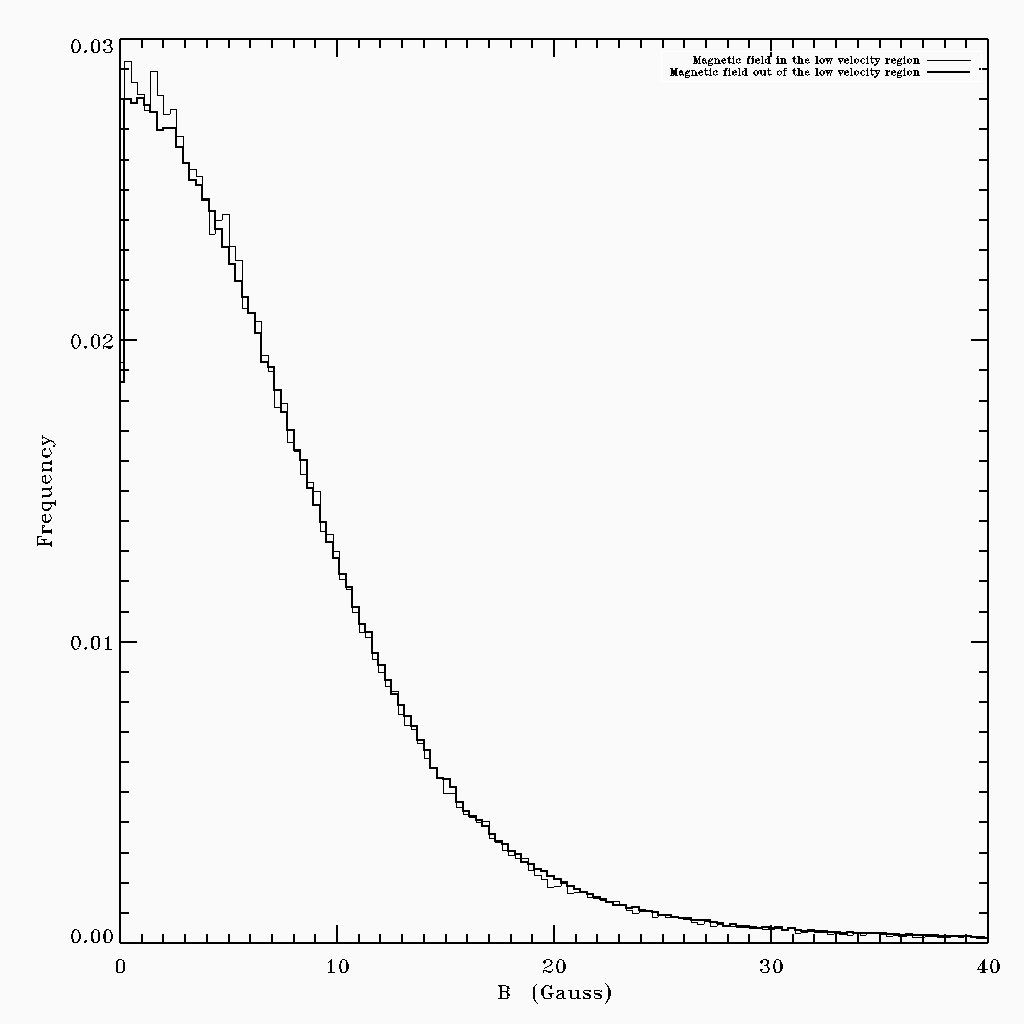}
\caption{Histograms of the absolute value of the magnetic flux in and out of the low velocity modulus region. The comparison of the histograms shows  a slightly higher magnetic flux amplitude in the lower velocity modulus region.}
\label{fig:donuts40}
\end{figure}

 A closer inspection of the texture of the velocity modulus map allowed us to locate  regions where donuts are almost absent or, at best, have low amplitudes compared to the rest of the solar surface.
 The longitudinal magnetic flux of such a region is shown in Fig.~\ref{fig:donuts39}. The comparison of the histograms of the absolute value of magnetic flux inside the lower velocity amplitude region  (in the white rectangle) and the rest of the magnetic flux  is plotted in Fig.~\ref{fig:donuts40}. A comparison of the histograms reveals  a slightly higher magnetic flux amplitude in the lower velocity modulus region, which is in agreement with the action of the magnetic field on the turbulent motions. However, the difference is small and we observed, in our data, only a few cases; this will have to be confirmed with a more statistical analysis.

\section{ Possible link between donuts  and dynamical events  of  the solar outer solar 
  atmosphere }\label{ssec:density}

     The potential action of the photospheric motions linked to donuts  onto the dynamics of the outer solar 
  atmosphere, such as the triggering of filament eruptions or many other applications on phenomena where surface motions are involved, is particularly interesting to contemplate. The solar atmosphere,  from the photosphere to  the chromosphere and  toward the corona, is not static.

In the photosphere, the magnetic field is subject to diffusion due to supergranular flows and the large-scale motions of differential rotation and meridional circulation \citep{Fan2021}. The magnetic field elements, which are transported across the solar surface by the surface motions, cause shearing of the chromospheric and coronal magnetic field.  The dynamics of the photospheric plasma has consequences on the properties of the chromosphere and the corona, in particular via the magnetic field whose elements propagate the disturbances related to the turbulence motion  of the solar surface. The strong coupling between convection and magnetic field in the photosphere has a consequence on the magnetic structures at higher solar atmospheric layers. For example, the convection cells transport the magnetic field and form a magnetic polarity inversion line between them \citep{Roudier2018}.
A previous study shows the importance of the photospheric flows  for influencing the stability of the structures of the magnetic field in upper layers of the solar atmosphere \citep{Woll2020}, for example in the destabilation  and eruption of solar filaments.

Other examples are described in the literature such as the  coronal plumes which are linked to photospheric flows. The bright plumes form where converging supergranular flows bring unipolar network elements together to form large, dense clumps \citep{Wang2016}. The plume emission fades when the flows diverge again and the clumps are dispersed \citep{Wang2016}.

The long temporal space-borne sequences' observations allow us to show the existence of long, persistent downflows together with the
magnetic field. They penetrate inside the Sun but are also connected with the anchoring of coronal loops in the photosphere, indicating
a link between downflows and coronal activity. A link suggests that EUV cyclones over the quiet Sun could be an effective way
to heat the corona  \citep{Roud2021}.

 \subsection{ Detection of donuts using properties of the divergence field}

 Different physical dynamical phenomena are observed on the Sun atmosphere, such as a vortex, a jet, blinkers,  coronal bright points (campfires), filaments' eruption, sunspot destruction, among others.   All of these phenomena occur in a temporal range from a few hours to several days and a spatial range from the solar granulation (1000 km) to scales of several megameters.
   As the donuts represent an intermediate scale of the described surface phenomena, we found it very interesting to explore potential links between these phenomena and donuts' evolutions.  Such studies require a more automatic and rapid detection of donuts. For this purpose, we developed a   divergence detection method, described below,  and we applied it to the new HMI/SDO sequences:  on 14 April 2016 containing plage regions, and on 26 January 2018 including an erupting filament.

   The detection of donuts by eye is limited since it is time-consuming and tedious. Nevertheless, this manual donut detection, described above, provided a sufficient sample  (554 donuts) to serve as a reference to implement new donut-detection methods  as automatically as possible.   From the analysis of  the 554 donuts, the donuts are found to represent the most powerful divergent events on the solar surface  related to supergranulation.  So, it was  natural to use the divergence field to automate the donut detection to get more detailed and statistical knowledge of their properties (shapes, amplitude distribution, etc.). Even if this method is incomplete, it must at least  allow us to detect the largest donuts in shape and amplitude with a good confidence level. The best method, we found,  is to use the divergence field calculated by the CST averaged over a few hours (12h or 24h).  The best method found is to apply a threshold to the divergence field and keep only the divergence with an amplitude above the selected threshold. 
  This gives a separated area, allowing for the largest amplitude divergence 
 in the field to be located.  The center of gravity of these  area elements allowed us to detect the  potential donut location.  

 The divergence threshold of $7.2 \times 10^{-5}$\,s$^{-1}$ allowed us to detect 95\% of donuts at the latitude and longitude $\pm20^\circ$ 
  with a great confidence.  Out of a total of 104 donuts detected visually, 99 donuts over 24h were well detected giving a detection error around 5\%. Considering the different amplitudes, shapes which are more or less closed, and the proximity of the donuts, that error is acceptable in a lot of studies such as those presented below.  
    The donut rate  found is  $1.7 \times 10^{-9}$ Mm$^{-2}$\,s$^{-1}$, which is close to those found for a larger field of view  ($1.05\times 10^{-9}$ Mm$^{-2}$\,s$^{-1}$) at the latitude and longitude of $\pm60^\circ$ with the manual method.

\begin{figure*}[!htbp]
\includegraphics[width=\hsize]{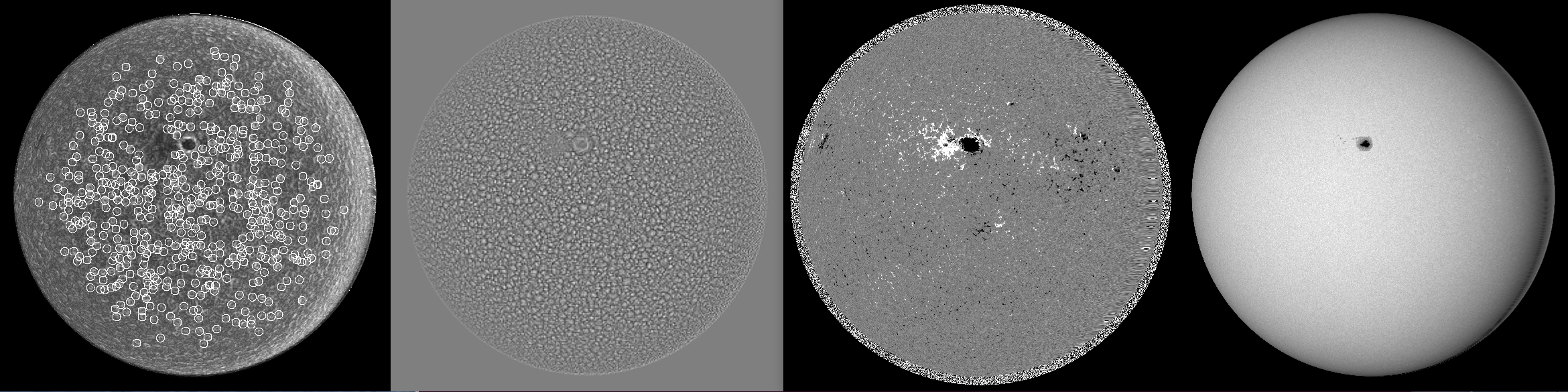}
\caption{Images of the Sun on 14 April 2016 showing spots and plages. From left to right: Velocity modulus $u_h$  and the detected donuts (circles), divergence  $\nabla\cdot\vec{u}_h$, longitudinal magnetic field, and continuum intensity.}
\label{fig:donuts43}
\end{figure*}

\begin{figure*}[!htbp]
\includegraphics[width=\hsize]{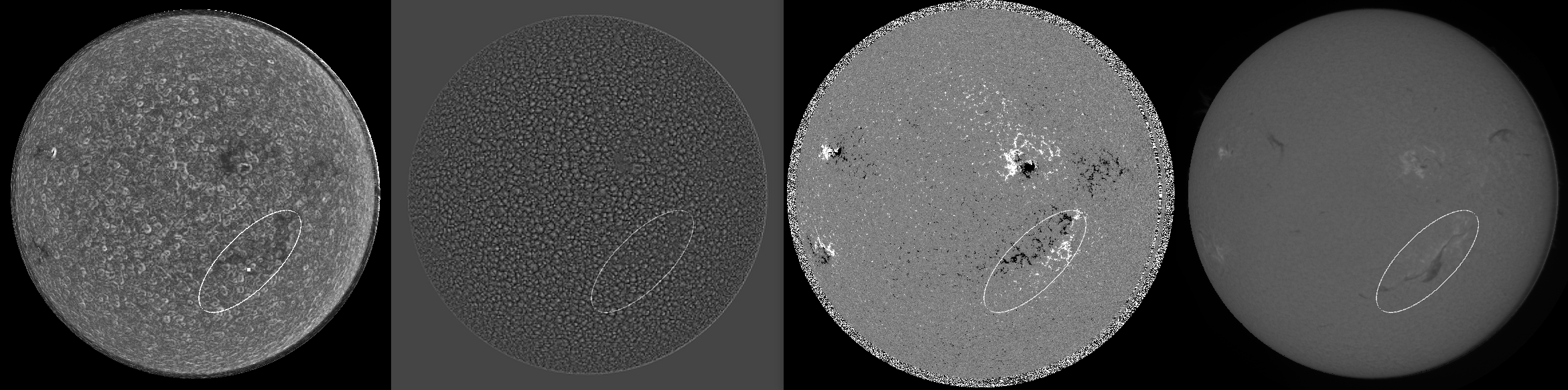}
\caption{Sunspot, plage, and filament on  26 January 2016. From left to right: Velocity modulus of $u_h$, divergence  $\nabla\cdot\vec{u_{h}}$ , longitudinal magnetic field, and H$\alpha$ observation (the ellipse indicates the filament region).} 
\label{fig:donuts44}
\end{figure*}

 \begin{figure*}[!htbp]
\includegraphics[width=\hsize]{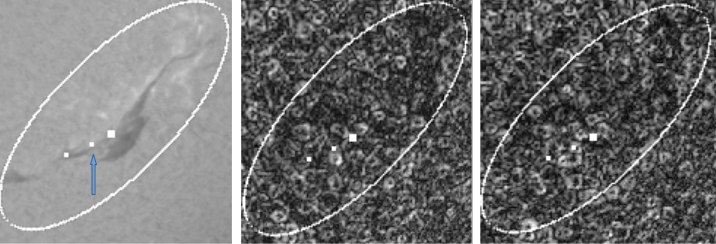}
\caption{Filament eruption in the H$\alpha$ line, on the left, on 26 January 2016 (16:30). The  square, indicated by the arrow, indicates the foot where the eruption started at 17:00. The middle and right images represent the horizontal velocity modulus at time 12:00 and 16:00, respectively. For both images, intense donuts are observed close to 12:00 and at the location foot eruption at 16:00. The field of view is $561\arcsec \times 572\arcsec$ ($401\times 408\:\mathrm{Mm}^2$)}
\label{fig:donuts70}
\end{figure*}

 \subsection{ Donuts and plage regions}\label{ssec:density2}

The  divergence detection method described in the previous section was applied to a new data set on 14 April 2016 and the results are shown in  Fig.~\ref{fig:donuts43}. The donuts are fairly evenly distributed on the solar surface, except in the magnetic regions (plages and sunspots). In particular, no donuts were detected in the plage region. The divergence is less structured and shows low amplitudes in plages and active regions in the north of the Sun. At the same location, we also observed a lower amplitude  of the velocity modulus and almost no donut. In the other western region with high amplitude magnetic fields, we also observed a lower velocity modulus and donuts are less present. This clearly indicates that the large magnetic field affects the supergranulation flows and, in particular, those with an axisymmetric shape, which are the most powerful supergranules.
It has already been established that in plage regions, which have higher magnetic fields than quiet regions, horizontal flows  are weaker by a factor of two   \citep{T92}.  We also know  that supergranulation is affected by the variation of the magnetic field. 
Our observation is compatible with the results found by  \citet{Meunier2007},  where a high level of weak fields may prevent the formation of large cells. Supergranules are smaller, on average, at cycle maximum than at cycle minimum \citep{Meunier2008}. 
    From our direct observations, we conclude that, in the magnetized areas, the strongest supergranular flows (donuts) are affected and thus also modify the diffusion of the magnetic field in these places.

\subsection{ Donuts and filament eruptions}\label{ssec:density3}

   From another observation of active area \citep[on 26 January 2016][]{Roudier2018}, we confirm the effects of the magnetic field on the formation of donuts (Fig.~\ref{fig:donuts44}). As described by \citet{Roudier2018}, in the neutral line (southwest) where a filament is located, we observed 
   weaker horizontal velocities, as well as the formation of donuts. 
    Before the filament eruption, we noted horizontal flow activity and intense donuts close to the location of one foot where the eruption started (close to the point C in Fig. 5 of \citet{Roudier2018}). This point, indicated by the arrow in Fig.~\ref{fig:donuts70}, is located close to large donuts before the eruption which could be one of the actors of the eruption triggering (Fig.~\ref{fig:donuts70}, middle and right). 
    However, it would be hazardous to draw a firm conclusion about the relation between filament destabilization and intense horizontal flows produced by donuts  from only one observation. A statistical study is needed to clarify the causal relationship between these intense flows and the eruption of the filament.

\subsection{ Potential link between donuts and the vortex}\label{ssec:futur}
\begin{figure*}[!htbp]
\includegraphics[width=\hsize]{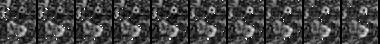}
\caption{Evolution of the velocity modulus $u_h$ on 29 November 2018. The time step between each frame is 1 h and the field of view spans  $127\arcsec \times 155\arcsec$ ($90.9\times 111\:\mathrm{Mm}^2$). White squares indicate points where the vorticity modulus has an amplitude greater than $5.5  \times 10^{-5}  \mathrm{s}^{-1} $.} 
\label{fig:donuts72}
\end{figure*}
        We now want to locate where the supergranulation is most effective and can be linked to local events. By studying previous  publications on supergranules and phenomena close to them (vortices, jets, etc.)           described in the literature in  detail, we notice that the studied supergranules were well structured, especially with  an isotropic distribution of their horizontal velocities as our donuts.
            We can cite, for example, the first work of \citet{ATTIE2009}, where we can see in their Figs. 3 and 7 the presence of a well-formed supergranule and the presence of a vortex.  
           
           Similarly, several analyses of an observation sequence of 2 November 2010 report that the supergranule is  well formed: readers can refer to Figure 1 of  \citet{GSBDB2014}, 
for example, where the horizontal velocity modulus is clearly isotropic as our donuts. This data set (2 November 2010) used in this work was exhaustively described  in \citet{GOS2016}, and previously analysed by \citet{GSBDB2014,GBBDS2014}  and \citet{ Caroli2015,GBC2018}.  \citet{Attie2016}  (Fig. 6 bottom)  observed  persisting structures such as large-scale vortex flows  ($\sim$18 Mm). This is confirmed by \citet{Reque2018} who, on data from 2-3 November 2010, report a detection of a similar vortex flow at supergranular junctions which consisted of three recurrent vortices (with an amplitude of $1.5 \times 10^{-4}  s^{-1} $) lasting for $\sim$7h  each and having diameters of $\sim$5 Mm.  \citet{Chian2020} observed  that persistent  vortices are formed in the gap regions of Lagrangian chaotic saddles at supergranular junctions with lifetimes varying from 28.5 to 298.3 min. 
             In the same way,   \citet{Roud2021} (Figs. 16,17,18) detected vorticity resulting from the large horizontal velocity amplitude of a supergranule with a donut shape and the combination of the velocities from the large neighboring supergranule.

           These photospheric vortices are probably linked to the whirling motions  discovered by   \citet{Wed2012} in the chromosphere, the atmospheric layer caught  between the photosphere and the corona. These structures are identified as observational signatures of rapidly rotating magnetic structures, which provide a mechanism for channeling energy from the lower into the upper solar atmosphere.

             In addition to the vortex flow close to the supergranule, different phenomena (jet, blinkers, coronal bright points, etc.) are reported in the literature. \citet{Muglach2021} describe coronal jets at the  photospheric footpoints  (Figs 1, 5 and 8)  located at the supergranular boundaries with a convergence of the horizontal flows.

               More recently, \citet{Pane2021} have  shown a link between small-scale transient brightenings (campfires) in the quiet Sun,  which occur at the edges of photospheric magnetic network flux lanes, similarly  to previous works \citep{Porter1987,Porter1991,Falc1998,Gosic2014,Attie2016}  in which active events are located at the boundaries of supergranule cells.   In the same way, in \citet{SHO2022}  the small-scale brightenings in the transition region (blinkers) are observed at the boundaries of supergranules (see their Fig. 8).
        
            All the phenomena presented here according to their description through the solar atmosphere are closely related to the supergranular flows and certainly the most powerful of them. As mentioned above, these observations are almost always related to very well-formed and generally clearly isotropic supergranular flows corresponding to the descriptions of our donuts.  The detection of the donuts, and particularly the larger ones, could be a way to locate 
            vortexes in the quiet Sun quickly.

             For this purpose, we have tested the localization of vorticity from our surface velocities.  From  ($u_\theta$,$u_\phi$), the vorticity fields have been computed with spatial (2.5 Mm) and temporal (30 minutes) steps, then averaged over 12~h with a running window. Figure~\ref{fig:donuts72} shows an example of the evolution of the large-amplitude vorticity region (greater than $5.5 \times 10^{-5}  s^{-1} $)) plotted over  the velocity modulus on 29 November 2018. In the bottom right of that figure, we observe long-lasting vorticity, around the large donuts, indicating a way 
             to locate the important vortex on the solar surface. However, in order to be more conclusive, a larger sample size is needed to be statistically significant and also with the simultaneous localization of magnetic and vorticity phenomena in the whole solar atmosphere. These results indicate the potentiality of the vortex  detection close to well-formed donuts.

\section{Conclusion}\label{sec:conc}

     On the Sun, the surface flows  guide the evolution of the magnetic field  elements at all scales of space and time. Supergranular diffusion and meridional flow advection of small-scale field elements play a role in the reversal of the global dipolar field.
     
    In this study, we are interested in looking at the texture of  the photospheric flows and the  relation between the surface velocity amplitude and the photospheric magnetic field. The  $u\theta~ $and$~ u\phi$ horizontal flows on the solar surface  issued from the CST 
     allowed us to follow these flows on the whole surface of the Sun with very high spatial (2.5 Mm) and temporal resolution (30 minutes).
    
     To characterize the kind of flow texture, the horizontal velocity modulus $u_h$  has been studied. That quantity shows  a different structuring from the divergence field issued from the same velocities, with different types of the spatial organization of the amplitude and shapes. In  $u_h$
maps, we observed  closed annular flows, not necessarily joint,  visible 
at all the latitudes which we referred to as donuts. However, the solar surface shows large areas without such elements with no structured flows, with a lower amplitude. These donuts with an isotropic distribution of the velocity modulus reflect the most active convective cells associated with supergranulation. Our measurements show that the donuts have identical properties (amplitude, shape, orientation) regardless of their position on the Sun.  Average  donuts computed from all the donuts at the disk center ($\pm$ 10\textdegree) shows an east-west asymmetry in the amplitude which is related to previous works on the wave-like properties of supergranulation \citep{LBG2018}. 
       
      A kinematic simulation of the donuts' outflow  represented by Gaussian sources, taking into account their locations, sizes, amplitudes, birth, and deaths, has been developed. That temporal simulated flows applied during 48h to passive particles (corks) show an empty area estimated to be 65\% of the total area at the disk center which mimics the photospheric network. This indicates the preponderant action of the selected donuts which are, from our analysis, one of the major actors for the magnetic field diffusion on the quiet Sun.
    
       Regarding the effect of the magnetic field on donuts formation, we observed the absence of donuts in the magnetized areas. In particular, the magnetic plages affect the strongest supergranular flows and thus this clearly modifies the diffusion of the magnetic field in these places.  From filament destabilization observation, we noted horizontal flow activity via intense donuts close to the location of one filament foot where the eruption started. Such results must be confirmed with more statistical material.
   
       One of the advantages of flow field  measurement by the CST  on long time series provided by SDO  is the detection of the isotropic supergranules and all the physical events associated all over the solar surface. First, SDO data generalize the HINODE observations at the disk center  to the whole surface of the Sun. The link between the families of granules and the  isotropic horizontal flows \citep{Roudier2016}  indicates that TFGs work underneath in a continuous way and could be one of the main actors of the photospheric surface flows. That scenario has to be confirmed with more new observations.
         
More generally, the different  phenomena observed in the solar atmosphere, such as blinkers,  coronal bright points (campfires), or vortexes, are also closely related to the most powerful supergranular flows, which correspond to our
         donut. Then, the detection of donuts, especially the largest ones, could be an interesting way to locate and study the evolution of the magnetic network, the destruction of sunspots, the eruption of filaments, and also their influence on the upper layers via spicules, jets, and vortexes.

 \begin{acknowledgements}
The HMI data is courtesy of the SDO HMI Science Investigation Team. Some of the computing was performed at the JSOC, which is located at the Stanford University and operated by the HMI Science Investigation Team. This work was granted access to the HPC resources of CALMIP under the allocation 2011-[P1115]. Thanks to C. Barnetche for her contribution to the neural networks processing tests. We appreciate numerous team members who have contributed to the success of the SDO and Hinode mission. The MEDOC data and operations centre (CNES / CNRS / Univ. Paris-Saclay) make  code availability and the CST documentation. These works are supported by COFFIES, NASA Grant 80NSSC20K0602. We thank the anonymous referee for her/his insightful comments, which have improved
the paper.
\end{acknowledgements}

\bibliographystyle{aa}    
\bibliography{biblio}

\appendix

\section{Movies}

 \begin{figure}[!htbp]
\includegraphics[width=\hsize]{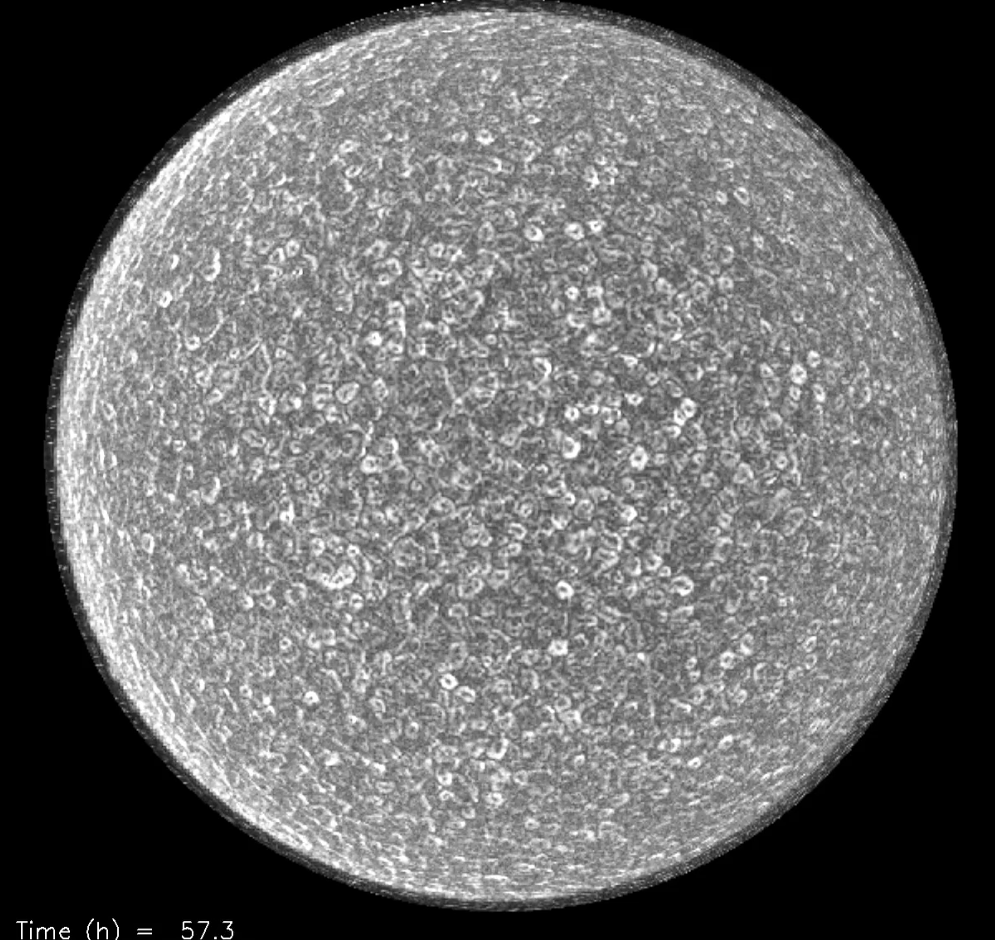}
\caption{(online movie)  Photospheric horizontal velocity modulus during 6 days, derotated from the solar differential rotation. A smoothing window of 12h was applied to reduce the noise. The reference day, center of the Sun, and solar radius for CST code were taken at 00:00  UTC of the fourth observation day, that is 29  November 2018.}
\label{fig:movie1}
\end{figure}
 
 \begin{figure}[!htbp]
\includegraphics[width=\hsize]{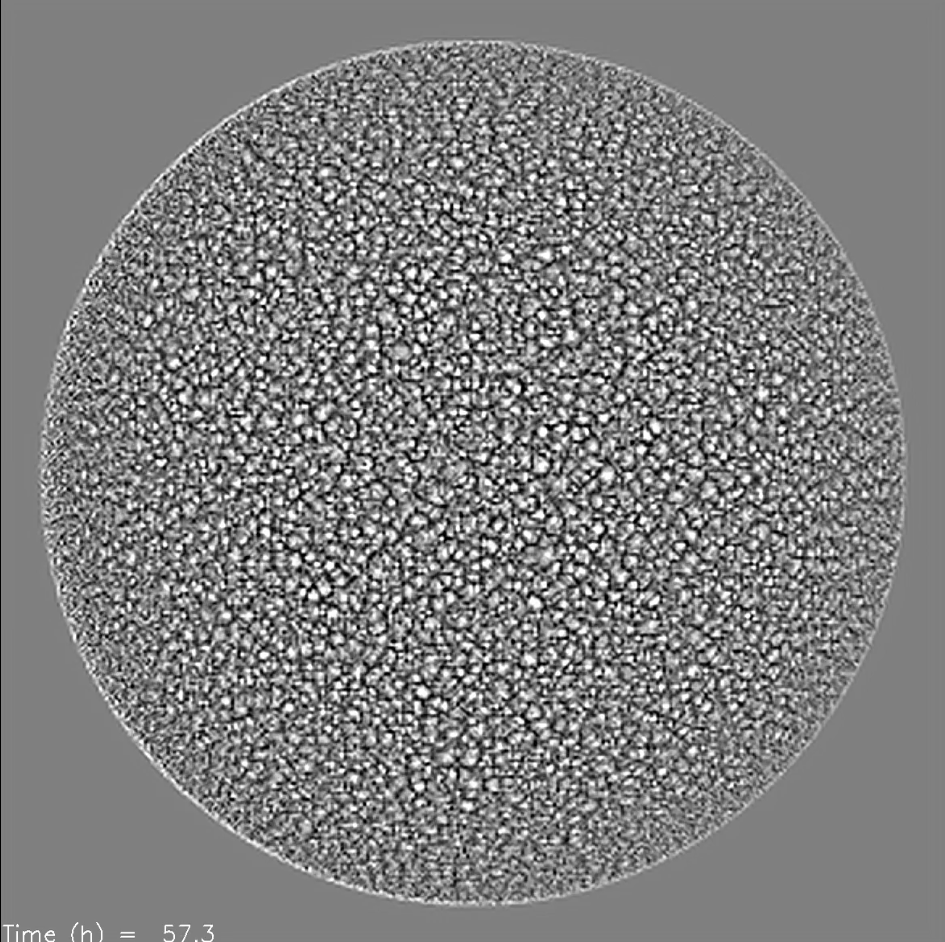}
\caption{(online movie) Divergence field computed from the horizontal velocity  during 6 days, derotated from the solar differential rotation. A smoothing window of 12h was applied to reduce the noise. The reference day, center of the Sun, and solar radius for CST code were taken at 00:00  UTC of the fourth observation day, that is 29  November 2018.}
\label{fig:movie2}
\end{figure}

 \begin{figure}[!htbp]
\includegraphics[width=\hsize]{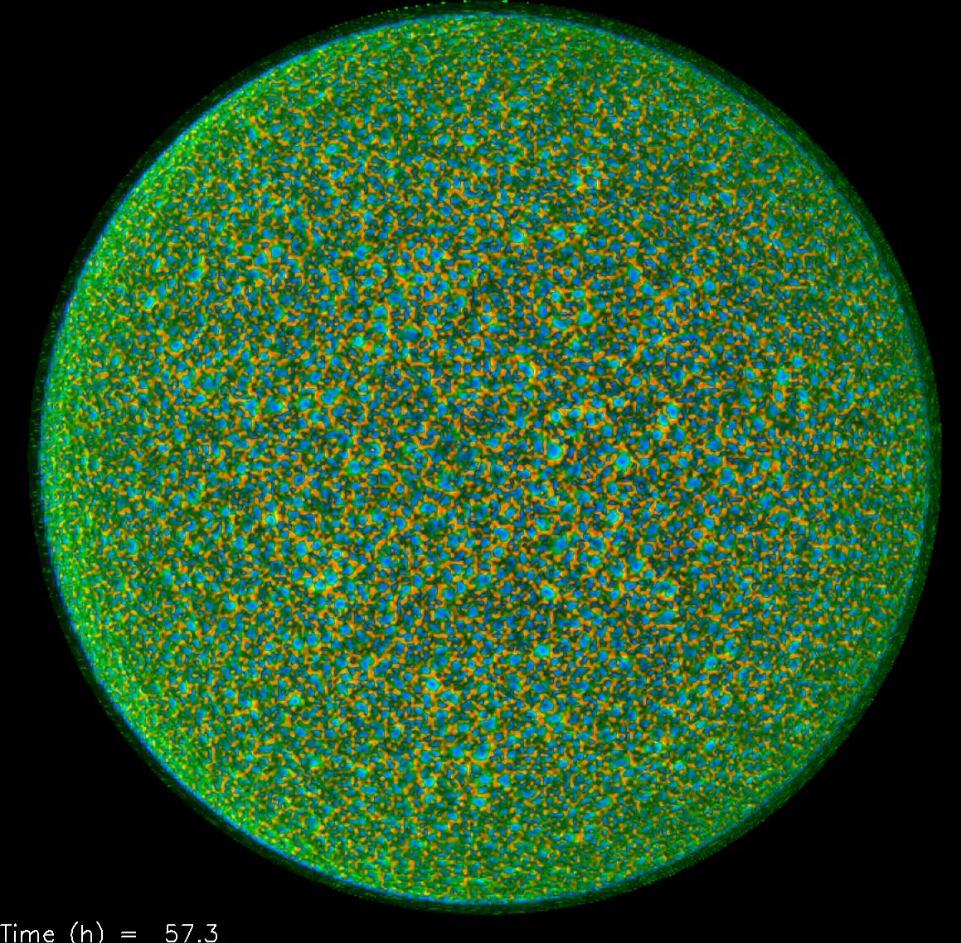}
\caption{(online movie) Combination of movie 1's horizontal velocity modulus and that of movie 2. 
  The horizontal velocity modulus was coded in white between 0 and 0.75 km/s, and the
  divergence in  red (convergence between 0 and $-1 \times 10^{-5}  \mathrm{s}^{-1}$) and blue (divergence between 0 and $+1 \times 10^{-5}  \mathrm{s}^{-1}$). We observed the maximum velocities surrounding the diverging areas.
}
\label{fig:movie3}
\end{figure}

\end{document}